\documentclass[twocolumn,tighten,times,usenatbib,twocolappendix,numberedappendix]{aastex631}

\usepackage{amsmath}
\usepackage{float}
\usepackage{color}
\usepackage{url}
\usepackage{stmaryrd}

\usepackage[percent]{overpic}

\usepackage{etoolbox}

\usepackage{amsmath}
\usepackage{amssymb}
\usepackage{graphicx}
\usepackage[flushleft]{threeparttable}
\usepackage{placeins}
\usepackage{xcolor}

\usepackage{siunitx}

\newcommand{\eq}[1]{Equation~\eqref{#1}}

\def\Msun{\,\mathrm{M}_{\odot} }

\def\va{v_\mathrm{a}}
\def\vo{v_\mathrm{o}}
\def\vw{v_\mathrm{w}}
\def\vr{v_\mathrm{r}}
\def\w{_\mathrm{w}}

\def\etaBHL{\eta_\mathrm{BHL}}

\def\etaBHL{\eta_\mathrm{BHL}}   
\hypersetup{breaklinks=true,colorlinks=true,linkcolor=blue,citecolor=blue,filecolor=blue,urlcolor=blue}
\makeatletter
\patchcmd{\NAT@citex}
  {\@citea\NAT@hyper@{%
     \NAT@nmfmt{\NAT@nm}%
     \hyper@natlinkbreak{\NAT@aysep\NAT@spacechar}{\@citeb\@extra@b@citeb}%
     \NAT@date}}
  {\@citea\NAT@nmfmt{\NAT@nm}%
   \NAT@aysep\NAT@spacechar\NAT@hyper@{\NAT@date}}{}{}

\patchcmd{\NAT@citex}
  {\@citea\NAT@hyper@{%
     \NAT@nmfmt{\NAT@nm}%
     \hyper@natlinkbreak{\NAT@spacechar\NAT@@open\if*#1*\else#1\NAT@spacechar\fi}%
       {\@citeb\@extra@b@citeb}%
     \NAT@date}}
  {\@citea\NAT@nmfmt{\NAT@nm}%
   \NAT@spacechar\NAT@@open\if*#1*\else#1\NAT@spacechar\fi\NAT@hyper@{\NAT@date}}
  {}{}

\usepackage{array}
\newcolumntype{C}[1]{>{\centering\let\newline\\\arraybackslash\hspace{0pt}}m{#1}}

\usepackage{xspace}

\def\apjl{ApJ}
\def\apjs{ApJS}
\def\ao{Appl.~Opt.}

\def\aap{A\&A}

\def\arcdeg{\hbox{$^\circ$}}

\definecolor{burgundy}{rgb}{0.5, 0.0, 0.13}

\usepackage{xspace}
\usepackage{wasysym}

\newcommand{\orcidicon}{\includegraphics[width=0.26cm]{orcid-ID.eps}}

\newcommand{\orcidauthor}[1]{\href{https://orcid.org/#1}{\orcidicon}}

\shorttitle{Wind accretion in binary systems}
\shortauthors{Emilio~Tejeda \& Jes\'{u}s~A.~Toal\'{a}}

\makeatletter
\patchcmd{\frontmatter@RRAP@format}{(}{}{}{}
\patchcmd{\frontmatter@RRAP@format}{)}{}{}{}
\renewcommand\Dated@name{}
\makeatother

\begin{document}

\title{\large Geometric correction for wind accretion in binary systems}

\correspondingauthor{Emilio Tejeda}
\email{emilio.tejeda@umich.mx}

\author[0000-0001-9936-6165]{Emilio Tejeda}
\affiliation{CONAHCyT - Instituto de F\'{i}sica y Matem\'{a}ticas, Universidad Michoacana de San Nicol\'{a}s de Hidalgo, Ciudad Universitaria, 58040 Morelia, Mich., Mexico}

\author[0000-0002-5406-0813]{Jes\'{u}s~A.~Toal\'{a}}
\affiliation{Instituto de Radioastronom\'{i}a y Astrof\'{i}sica, Universidad Nacional Aut\'{o}noma de M\'{e}xico, Morelia 58089, Mich., Mexico}

\date[]{Accepted to ApJ}

\begin{abstract}
\noindent 
The Bondi-Hoyle-Lyttleton (BHL) accretion model is widely used to describe how a compact object accretes material from a companion's stellar wind in binary systems. However, its standard implementation becomes inaccurate when the wind velocity ($\vw$) is comparable to or less than the orbital velocity ($\vo$), predicting non-physical accretion efficiencies above unity. This limits its applicability to systems with low wind-to-orbital velocity ratios ($w=\vw/\vo \leq 1$), such as symbiotic systems. We revisit the implementation of the BHL model and introduce a geometric correction factor that accounts for the varying orientation of the accretion cylinder relative to the wind direction. This correction ensures physically plausible accretion efficiencies ($\eta\leq 1$) for all $w$ in circular orbits. Our new implementation naturally predicts the flattening of the accretion efficiency observed in numerical simulations for $w<1$, without the need for {\it ad hoc} adjustments. We also peer into the implications of our prescription for the less-explored case of eccentric orbits, highlighting the key role of the geometric correction factor in shaping the accretion process. We compare our predictions with numerical simulations, finding good agreement for a wide range of parameters. Applications to the symbiotic star R~Aqr and the X-ray binary LS 5039 are presented. This improved implementation offers a more accurate description of wind accretion in binary systems, with implications for stellar evolution, population synthesis, and observational data interpretation.
\end{abstract}

\keywords{
\href{http://astrothesaurus.org/uat/154}{Binary Stars~(154)}; 
\href{http://astrothesaurus.org/uat/1578}{Stellar accretion~(1578)};
\href{http://astrothesaurus.org/uat/1636}{Stellar Winds~(1636)};
\href{http://astrothesaurus.org/uat/733}{High mass X-ray binary stars~(733)};
\href{http://astrothesaurus.org/uat/1674}{Symbiotic binary stars~(1674)}
}

\section{Introduction}
\label{sec:intro}

\label{sec:intro}

Mass transfer in binary systems plays a crucial role in understanding accretion onto compact objects and stellar evolution. This process is key in a wide range of scenarios, powering phenomena such as cataclysmic and symbiotic binaries, novae, supernovae, and X-ray binaries \citep{Frank2002}. 
The Bondi-Hoyle-Lyttleton (BHL) accretion model, developed initially by \citet{HL39} and later refined by \citet{BH44}, constitutes a fundamental framework for analyzing this mass transfer by providing a theoretical estimate of the rate at which a compact object accretes material from a wind. 

The BHL model's strength lies in its straightforward approach, offering a clear basis for understanding wind accretion. Its relative simplicity makes it a valuable tool, serving as a starting point for more complex accretion scenarios \citep{RomeroVila}. However, it is important to recognize that the model's underlying assumptions, such as a point-like accretor and a uniform supersonic wind, simplify the complexities of real astrophysical systems. Factors like the non-uniform nature of stellar winds, interactions with radiation fields, and the influence of magnetic fields necessitate modifications and extensions to the original BHL formalism \citep{Edgar04}.

The first application of the BHL model to the specific case of accreting binaries was presented by \citet{Davidson73}, who used it to model the X-ray sources Cen X-3 and Her X-1. In these systems, a magnetized neutron star accretes material from the wind of a massive companion star. Throughout the present work, we will frequently refer to the specific implementation of the BHL model presented  by \citet{Davidson73} as the ``standard BHL implementation.'' This is done for brevity and clarity, as their approach has become the most widely used and recognized application of the BHL model in the context of accreting binaries \citep{Boffin1988, Theuns96,nagae04,Liu17,Saladino19,ElMellah19}.

The BHL model has proven to be a versatile tool for understanding a wide range of systems. In the context of high-mass X-ray binaries (HMXBs), the standard BHL implementation provides a basic understanding of the accretion process, the resulting X-ray emission \citep{Davidson73,Okuda77,Reig03,Negueruela10,Ducci10,ReviewHMXRB}, and the estimation of key system parameters \citep{Eadie75}. Beyond  HMXBs, the standard BHL implementation has also been employed in the study of symbiotic binaries, where a white dwarf (WD) accretes from the wind of a late-type giant companion, providing insights into the accretion rates that can trigger outbursts and influence stellar evolution \citep{ValBorro09,Chen18,Saladino18,Vathachira2024}. Moreover, it has been also used as benchmark for the more complex common envelope scenario \citep[e.g.,][]{MacLeod15,LC22,ElBadry23,Moreno23}.

The BHL model is applicable to supersonic flows, requiring the relative velocity between the gas and the accretor ($\vr$) to exceed the local speed of sound ($c_\mathrm{s}$). In a binary, this translates to $c_\mathrm{s} < \vr \simeq \sqrt{\vw^2 + \vo^2}$, where $\vw$ is the wind velocity at the position of the accretor and $\vo$ is the orbital velocity. This condition is readily met in HMXBs, where the winds of massive OB stars quickly reach supersonic speeds close to the stellar radius \citep{Castor1975}. In contrast, for late-type stars, as those found in symbiotic systems, the dust-driven stellar winds reach supersonic speeds ($\vw > c_\mathrm{s}\approx 2$ km~s$^{-1}$) at up to few stellar radii \citep{ElMellah2020}. However, a compact companion would have a typical orbital speed of $v_\mathrm{o}\simeq 10-50\,\mathrm{km\,s}^{-1}$, ensuring  a supersonic relative velocity.

The standard implementation of the BHL model, while widely used in the context of wind-fed accreting binaries, faces limitations when the wind velocity $\vw$ is comparable to or less than the orbital velocity $\vo$ of the accretor \citep[][]{Boffin15,Hansen2016}. This is particularly relevant for symbiotic systems for which simulations predict a flattening of the mass accretion efficiency when $\vw \ll \vo$ \citep{nagae04,Saladino19}, while the formula for the accretion rate of the standard BHL implementation typically overestimates the accretion rate and, for sufficiently small wind-to-orbital velocity ratios, can lead to non-physical values exceeding unity. For instance, observations of the symbiotic system $o$ Ceti predict a mass accretion onto its WD component of 10$^{-10}$~M$_\odot$~yr$^{-1}$, a couple of orders of magnitude below the standard BHL prediction \citep[][]{Sokoloski2010}. Such discrepancies highlight the need for refinements, especially in scenarios with low wind-to-orbital velocity ratios. 

Several authors have proposed modifications to the standard BHL implementation to address these limitations, for example, incorporating arbitrary efficiency factors or cut-off values \citep[see e.g.,][]{nagae04,Saladino19,Malfait2024}. However, these modifications often have limited applicability. In this article, we revisit the implementation of the BHL model for accreting binaries and introduce a simple geometric correction factor that yields improved agreement with simulations across a wider range of wind-to-orbital velocity ratios. 

Notably, our modified implementation naturally reproduces the flattening of the mass accretion efficiency observed in simulations when  $\vw \ll \vo$, without the need for arbitrary adjustments. Furthermore, we extend our analysis to binary systems with eccentric orbits with the purpose of testing our predictions in these more intricate systems, which are crucial for understanding the long-term evolution and observational signatures of a variety of astrophysical phenomena.

This paper is organized as follows. Section~\ref{sec:themodel} describes our new prescription for implementing the BHL model in accreting binaries, detailing the geometric correction factor and the resulting mass accretion efficiency.  Section~\ref{sec:predictions} presents the results for both circular and elliptical orbits, including comparisons with numerical simulations. Section~\ref{sec:discussion} discusses the implications of these results, particularly for symbiotic stars and HMXBs, and addresses the limitations and potential refinements of the model. Finally, Section~\ref{sec:summary} summarizes the key findings.

\section{Analytical model}
\label{sec:themodel}

\begin{figure}
\begin{center}
\includegraphics[width=\linewidth]{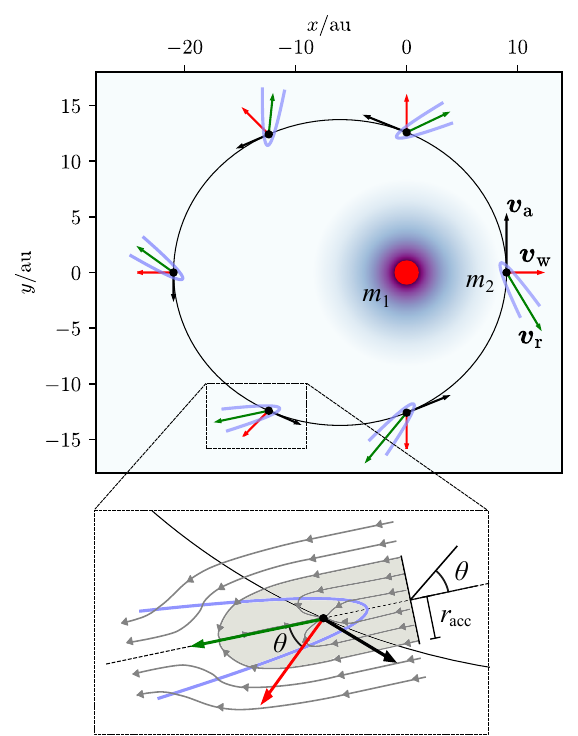}
\end{center}
\caption{Schematic representation of wind accretion in a binary system, highlighting the geometric correction factor. The binary system has $m_1 = 3\Msun$, $m_2 = 1\Msun$, $a = 15\,$ au, and $e = 0.4$, with a constant radial wind velocity of $20$ km\,s$^{-1}$. The accretor's elliptical trajectory around the primary is shown, with velocity vectors for the accretor ($\boldsymbol{v}_\mathrm{a}$, black arrow), the primary's stellar wind ($\boldsymbol{v}_\mathrm{w}$, red arrow), and their relative velocity ($\boldsymbol{v}_\mathrm{r}$, green arrow). Parabolic shapes qualitatively indicate the expected location of the accretion wake at each position. {\it Inset}: Accretion flow streamlines around the secondary. The shaded area represents the fraction of the incoming wind accreted by the secondary, visualized as being within an accretion cylinder of radius $r_{\rm acc}$. The angle $\theta$ between $\boldsymbol{v}_\mathrm{w}$ and $\boldsymbol{v}_\mathrm{r}$ highlights the need to account for a projection factor when applying the BHL model to binary systems.}
\label{diagram}
\end{figure}

We consider a binary system composed of a donor star and an accreting companion of masses $m_1$ and $m_2$, respectively.
The primary star loses mass via a stationary stellar wind, and the secondary accretes material from this wind as it goes around its trajectory (see Figure~\ref{diagram}). 

Given the orbital parameters, semi-major axis $a$ and eccentricity $e$, the accretor's trajectory around the primary is described by
\begin{gather}
    r(\varphi) = \frac{a(1-e^2)}{1+e\cos\varphi}, \\
    \boldsymbol{v}_a = \frac{v_\mathrm{o}}{\sqrt{1-e^2}}\left[ (e\sin\varphi)\,\hat{r} + (1+e\cos\varphi)\,\hat{\varphi}  \right],
\end{gather}
where $v_\mathrm{o} = \sqrt{G(m_1+m_2)/a}$ is the mean orbital velocity of the accretor, $\varphi$ is the true anomaly, and $\hat{r}$ and $\hat{\varphi}$ are the unit vectors in the radial and azimuthal directions, respectively. 

The stellar wind of the donor can be characterized by its mass-loss rate $\dot{M}_\mathrm{w}$. Assuming that the stellar wind is stationary and isotropic, at any distance $r$ from the primary, it satisfies
\begin{equation}
    \Dot{M}\w = 4\pi r^2 v\w \rho\w,
\end{equation}
where $v\w$ is the wind velocity and $\rho\w$ is the wind density at radius $r$. We adopt a $\beta$ velocity law for the wind \citep{Lamers1999}
\begin{equation}
        \boldsymbol{v}\w(r) = \left[ v_\mathrm{i} + (v_\infty - v_\mathrm{i}) \left(1 - \frac{R}{r}\right)^\beta \right]\hat{r},
        \label{eq:beta}
\end{equation}
\noindent where $v_\infty$ is the terminal wind speed, $R$ is the wind launching radius, $v_\mathrm{i}$ is the initial velocity at $R$, and $\beta$ is an exponent determining how quickly the terminal speed is reached. This analytical prescription, defined for $r \ge R$, provides a good approximation of the acceleration process found in line-driven winds (hot stars) and those driven by radiation pressure on dust grains (late-type stars).

The relative velocity between the accretor and the stellar wind is  
\begin{equation}
    \boldsymbol{v}_\mathrm{r} = \boldsymbol{v}_\mathrm{w} - \boldsymbol{v}_\mathrm{a}, 
\end{equation}
with its magnitude given by
\begin{equation}
    \vr = \sqrt{\vw^2 + \va^2 - 2\, \vw \va^r },
    \label{vr_ecc}
\end{equation}
where 
\begin{equation}
    \va = \vo\sqrt{\frac{2a}{r} - 1} 
\end{equation}
\noindent is the magnitude of the accretor's velocity and
\begin{equation}
    \va^r = \frac{e\, \vo  \sin\varphi}{\sqrt{1-e^2}},
\end{equation}
is its radial component.

According to the BHL model, a point mass accretes from a wind with which it has a relative supersonic velocity $\vr$ at a rate
\begin{equation}
    \Dot{M}_\text{BHL} = \pi r_\mathrm{acc}^2 \, \rho\w \,\vr, 
    \label{eq:BHL}
\end{equation}
where
\begin{equation}
    r_\mathrm{acc} = \frac{2Gm_2}{\vr^2},
    \label{racc}
\end{equation}
is the radius of the accretion cylinder as shown in the inset of Figure~\ref{diagram}.

To implement \eq{eq:BHL} for accreting binaries in elliptical orbits, we must account for the varying orientation of the accretion cylinder relative to the wind direction. Assuming a uniform and spherically symmetric wind, the fraction of wind intercepted by the accretion cylinder's cross-sectional area depends on its instantaneous orbital position. The captured flux is determined by the projected area of the accretion cylinder's cross-section onto a sphere centered around the primary. This projection accounts for the effective area capturing the wind. Since the relative velocity $\boldsymbol{v}_\mathrm{r}$ is perpendicular to this cross-section (see inset in Figure~\ref{diagram}), the angle $\theta$ between the wind velocity $\boldsymbol{v}\w$ and the accretion cylinder's cross-sectional area can be determined from
\begin{equation}
 \cos\theta  = \frac{\boldsymbol{v}_\mathrm{r} \cdot \boldsymbol{v}\w}{ \vr v\w } = \frac{v\w - \va^r}{\vr}  .
 \label{cos}
\end{equation}

The mass accretion efficiency ($\eta$) is defined as the fraction of the donor's total expelled mass flux ($\dot M_\mathrm{w}$) that is ultimately accreted by the secondary. Accounting for the geometric projection factor, this efficiency is given by
\begin{equation}
    \eta = \frac{\Dot{M}_\text{BHL}}{\dot M\w} \cos\theta 
       = \frac{1}{4}\left|1-\frac{\va^r}{\vw}\right| \bigg(\frac{r_\mathrm{acc}}{r}\bigg)^2.
       \label{eta}
\end{equation}

In contrast, neglecting the geometric projection factor leads to the result of the standard BHL implementation
\begin{equation}
    \etaBHL =  \frac{1}{4} \frac{\vr}{\vw} \bigg(\frac{r_\mathrm{acc}}{r}\bigg)^2 ,
    \label{etaBHL}
\end{equation}
which has been widely employed in the literature \citep[see, e.g.,][]{Boffin1988, Theuns96,nagae04,Liu17,Saladino19,ElMellah19,Vathachira2024}. Note that \eq{eta} reduces to the standard result in \eq{etaBHL} when the accretion cylinder is perfectly aligned with the wind direction ($\theta=0$). This condition is formally satisfied only in the limit where the wind velocity significantly exceeds the orbital velocity ($\vw \gg \vo$).

It is worth noting that expressions for the mass accretion efficiency similar to Equation~(\ref{eta}), albeit with different underlying assumptions, have been previously presented in the literature with less clear derivations \citep{Reig03,Bozzo21}. However,  by explicitly considering the geometry of the accretion flow, our formulation provides a more rigorous and transparent derivation of the mass accretion efficiency, leading to a clearer physical interpretation and broader applicability.

\subsection{Refill time}

An important assumption underlying the prescription in \eq{eta} is that the wind has sufficient time to return to an unperturbed state between consecutive passages of the accreting object.  The accretion wake in the vicinity of the accretor is characterized by a length scale of $r_\mathrm{acc}$. Therefore, we can estimate the refill time ($t_\mathrm{fill}$) as the time required for the wind, traveling at a speed of $\vw$, to cover this distance between the accretor's successive passages. In other words, the refill time can be expressed as
\begin{equation}
    t_\mathrm{fill} = \frac{r_\mathrm{acc}}{\vw} = \frac{2Gm_2}{\vw \vr^2}.
\end{equation}
For our prescription to be applicable, we require that this refill time is significantly shorter than the orbital period,  $T=2\pi G(m_1+m_2)/\vo^3$, which results in the following condition
\begin{equation}
   \zeta = \frac{t_\mathrm{fill}}{T} = \left(\frac{m_2}{m_1 + m_2}\right) \left( \frac{\vo^3}{\pi \vw \vr^2} \right) \ll 1 .
    \label{eq:l1}
\end{equation}

With the exception of a few specific cases detailed in Section~\ref{circ}, all scenarios presented in this work satisfy the aforementioned condition.

\section{Model predictions}
\label{sec:predictions}

\begin{figure}
\begin{center}
\includegraphics[width=\linewidth]{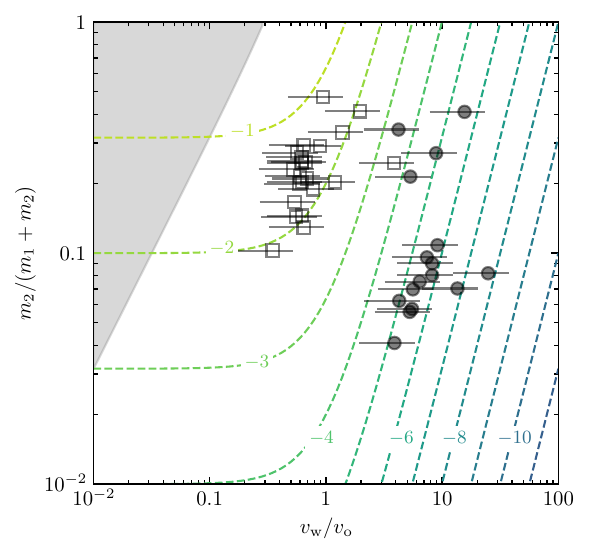}
\end{center}
\caption{Mass accretion efficiency $\eta$ contour plot for circular orbits (Equation~\ref{eta_c}) as a function of the dimensionless wind velocity $w = \vw/\vo$ and mass ratio $q = m_2/(m_1+m_2)$. Dashed lines indicate contours of constant $\log_{10}(\eta)$. The shaded area represents the region of the parameter space where the condition in \eq{eq:l1} is not met. This region is bounded by the solid gray line where \eq{eq:l1} reaches equality ($\zeta = 1$).
The parameter space occupied by observed high-mass X-ray binaries (HMXB) is represented by filled bullets, assuming a wind velocity range of 1000--3000 km~s$^{-1}$ with an average of 2000 km~s$^{-1}$. Empty squares represent symbiotic systems, where a wind velocity range of 10--30 km~s$^{-1}$ and an average of 20 km~s$^{-1}$ have been assumed. See Appendix~\ref{sec:obs} for further details on the data.} 
\label{fig:param_space}
\end{figure}

\subsection{Circular orbits}
\label{circ}

\begin{table*}
\begin{center}
\footnotesize
\caption{Comparison of the mass accretion efficiency ($\eta$) predicted by our modified BHL implementation with the standard expression ($\eta_\mathrm{BHL}$) and numerical simulations ($\eta_\mathrm{sim}$). For eccentric orbits, the reported efficiencies correspond to averaged values over a full orbital period (see Eq.~\ref{eq:averaged}).}
\begin{tabular}{lcccccccccccc}
\hline
Reference & $e$ & $m_1$ & $m_2$ & $q$ &$v_\mathrm{o}$ & $v\w \symbol{92} v_\infty^{\phantom{w}\star}$ & $w$ & $\beta$ & $v_\mathrm{i}$ & $\eta_\mathrm{sim}$ & $\eta$ & $\etaBHL$  \\
   & & [M$_\odot$] & [M$_\odot$] & & [km~s$^{-1}$] & [km~s$^{-1}$] &  &  & [km~s$^{-1}$]   & [\%] & [\%] & [\%] \\
\hline
\citet{Theuns96} & 0.0 & 3 & 1.5 & 0.33 & 36.5 & 15 & 0.42 & \dots & \dots & 8 &  8.1 & 21.4    \\
\citet{ValBorro09} & 0.0 & 1.2 & 0.6 & 0.33 & 10.00 & 6 & 0.60 & \dots & \dots & 6 &  6 & 11.6 \\
\cite{HE13} & 0.0 & 1.5 & 1.0 & 0.40 & 14.90 & 10 & 0.67 & \dots & \dots & 8.5 &  7.6 & 13.6 \\
            & 0.0 & 1.5 & 1.0 & 0.40 & 12.16 & 10 & 0.82 & \dots & \dots & 5 &  5.7 & 9.0 \\
            & 0.0 & 1.5 & 1.0 & 0.40 & 10.53 & 10 & 0.95 & \dots & \dots & 3 &  4.4 & 6.4 \\ 
\citet{Chen2020} & 0.0 & 1.02 & 0.51 & 0.33 & 14.40 & 15.8 & 1.10 & \dots & \dots & 1.8 &  2.3 & 5.3 \\ 
\citet{Lee2022}   & 0.0 & 1.5 & 0.6 & 0.29  & 30.50 & 13.7 & 0.45 & \dots & \dots  & 20.8$\pm$5.6 &  5.7 &  13.8 \\
                  & 0.0 & 1.5 & 0.6 & 0.29  & 30.50 & 15.1 & 0.50 & \dots & \dots  & 11.1$\pm$3.8  & 5.3 &  11.9 \\
                  & 0.0 & 1.5 & 0.6 & 0.29  & 30.50 & 16.7 & 0.55 & \dots & \dots & 12.4$\pm$3.6 &  4.8    &  10.1 \\
                  & 0.0 & 1.5 & 0.6 & 0.29  & 30.50 & 18.5 & 0.61 & \dots & \dots & 4.8$\pm$1.3 &  4.4 &   8.4 \\
                  & 0.0 & 1.5 & 0.6 & 0.29  & 30.50 & 20.4 & 0.67 & \dots & \dots & 3.6$\pm$2.2 &  3.9 &   7.0 \\
                  & 0.0 & 1.5 & 0.6 & 0.29  & 30.50 & 21.8 & 0.71 & \dots & \dots & 4.3$\pm$1.9 &  3.6 &   6.2 \\
                  & 0.0 & 1.5 & 0.6 & 0.29  & 30.50 & 23.9 & 0.74 & \dots & \dots & 2.2$\pm$1.9 &  3.1 &   5.1 \\
\citet{Malfait2024} & 0.0 & 1.5 & 1.0 & 0.40 & 19.23 & 12.4 & 0.64 & \dots & \dots & 12.5 &  8.0 & 14.7 \\
                    & 0.0 & 1.5 & 1.0 & 0.40 & 19.23 & 14.8 & 0.77 & \dots & \dots & 9.9  &  6.3 & 10.3\\
                    & 0.0 & 1.5 & 1.0 & 0.40 & 19.23 & 22.6 & 1.18 & \dots & \dots & 4.0  &  2.8 & 3.7\\
\hline
\citet{SaladinoPols2019} & 0.2 & 1.2 & 0.6 & 0.33 & 17.90 & 15.4 & \dots & 0.88 & 12 & 7.1 & 4.2 & 11.4 \\
                         & 0.4 & 1.2 & 0.6 & 0.33 & 15.51 & 15.4 & \dots & 0.88 & 12 & 4.6 & 3.3 &  8.2\\
                         & 0.6 & 1.2 & 0.6 & 0.33 & 12.65 & 15.4 & \dots & 0.88 & 12 & 2.6 & 2.4 &   5.4\\
                         & 0.8 & 1.2 & 0.6 & 0.33 & 8.95  & 15.4 & \dots & 0.88 & 12 & 1.0 & 1.2 &  3.0\\
\citet{Malfait2024} & 0.5 & 1.5 & 1.0 & 0.40  & 19.23 & 15.5 & \dots & 1.23 & 5  & 20.6 &  8.2 & 17.6 \\
                    & 0.5 & 1.5 & 1.0 & 0.40  & 19.23 & 17.7 & \dots & 1.94 & 10 & 12.3 &  6.7 & 13.0 \\
                    & 0.5 & 1.5 & 1.0 & 0.40  & 19.23 & 24.8 & \dots & 2.65 & 20 & 4.3  &  3.2 & 4.7 \\
\hline
\end{tabular}
\begin{tablenotes}
      \item $^{\star}$This column shows the wind velocity: $v_\mathrm{w}$ (unperturbed value at the accretor's location) for circular orbits, and $v_\infty$ (terminal velocity) for eccentric orbits.
\end{tablenotes}
\label{tab:comparison}
\end{center}
\end{table*}

For circular orbits, we can simplify the mass accretion efficiency expression given by Equation~\eqref{eta} as
\begin{equation}
    \eta = \left[\frac{G m_2}{a(\vo^2+\vw^2)}\right]^2 = \left(\frac{q}{1 + w^2}\right)^2,
    \label{eta_c}
\end{equation}
where we have introduced the dimensionless mass and velocity ratios
\begin{gather}
    q = \frac{m_2}{m_1 + m_2}, \label{eq:q}\\
    w = \frac{\vw}{\vo}.
\end{gather}
This result is consistent with that of \citet{Hiari2021}, who employed a different derivation. Interestingly, in the case of circular orbits, the expressions derived by \citet{Reig03} and \citet{Bozzo21}, despite their differing approaches, also reduce to Equation \eqref{eta_c}, yielding the same result as our model.

The concise formulation in \eq{eta_c} underscores that, for circular orbits, the mass accretion efficiency is entirely determined by the ratios $q$ and $w$.
We illustrate this dependence in Figure~\ref{fig:param_space}, plotting contours of $\eta$ as a function of $q$ and $w$. The shaded region represents the parameter space where the refill time equals or exceeds the orbital period, which violates Equation~(\ref{eq:l1}) and thus falls outside the scope of our model. To provide observational  context, we also included the parameter space occupied by representative symbiotic systems and HMXBs (see Appendix~\ref{sec:obs} for the complete list of objects). It is noteworthy that the four highest data points in the figure correspond to HMXBs known to host black holes.

From Equation~\eqref{eta_c}, we can derive the asymptotic behavior of the mass accretion efficiency $\eta$ in two limiting regimes. When the wind speed significantly dominates the orbital velocity ($w\gg1$), the efficiency scales as
\begin{equation}
    \eta \simeq \frac{q^2}{w^4},
    \label{eq:limit1}
\end{equation}
which recovers the behavior of $\etaBHL$ (Equation~\ref{etaBHL}) in this regime.
Conversely, when the orbital velocity is much larger than the wind speed ($w\ll1$), the efficiency becomes independent of $w$ and approaches a constant value 
\begin{equation}
    \eta \simeq q^2.
    \label{eq:limit2}
\end{equation}
This contrasts with $\etaBHL$, which grows without limit as $q^2/w$ in this regime.

A key advantage of our formulation in Equation~\eqref{eta_c} is that it guarantees physically plausible mass accretion efficiencies ($\eta \leq 1$) for any combination of values $q$ and $w$. This is especially relevant for low wind-to-orbital velocity ratios ($w \ll 1$), a regime where the standard BHL implementation leads to non-physical predictions, as discussed in Section \ref{sec:intro}. 

To assess the accuracy of our modified BHL implementation, we compare its predictions to those from existing numerical simulations. We selected representative works from the literature, encompassing a range of relevant parameters and physical conditions, including variations in the primary's wind acceleration and numerical methods employed. These simulations provide values for the mass accretion efficiency ($\eta_\mathrm{sim}$), which we compare with our new prescription for $\eta$ (Equation \ref{eta_c}) and the standard expression $\eta_\mathrm{BHL}$ (Equation \ref{etaBHL}). A summary of this comparison is shown in Table~\ref{tab:comparison}, demonstrating strong overall agreement between our modified analytical estimate of $\eta$ and $\eta_\mathrm{sim}$ from different works.

We note that the largest discrepancies between our model and the simulations are found in the studies by \citet{Lee2022} and \citet{Malfait2024}, where the wind is still being accelerated within the spatial domain of interest, particularly for the cases with the smallest wind velocities. This ongoing acceleration complicates the comparison, as it becomes ambiguous to choose an appropriate value for the wind velocity in our model, which ideally should correspond to the unperturbed wind velocity at the location of the secondary. We will explore the possible implications of this ambiguity and other factors contributing to the discrepancies in Section \ref{sec:discussion}.

\begin{figure}
\begin{center}
\includegraphics[width=\linewidth]{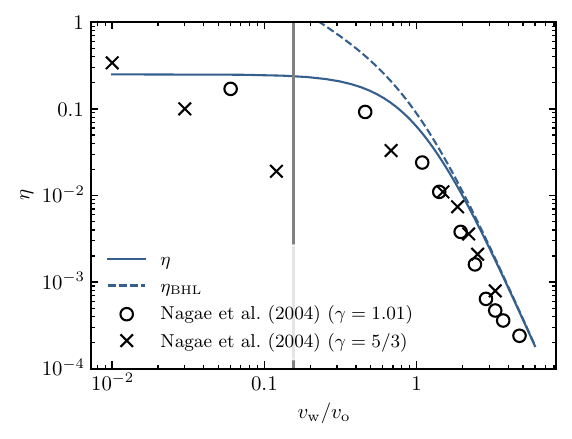}
\end{center}
\caption{Comparison between the mass accretion efficiency predicted by the modified BHL implementation proposed in this work ($\eta$ -- solid line) versus the standard one ($\eta_\mathrm{BHL}$ -- dashed line). The calculations have been computed for a binary system with $q=0.5$. The symbols show the results from numerical simulation from \citet{nagae04}. The vertical line at $w \simeq 0.15$ marks the point where the condition in \eq{eq:l1} is no longer met. Points to the left of this threshold, where the refill time is larger than the orbital period, should in principle lay outside the scope of our model.}
\label{fig:fig_nagae}
\end{figure}

In Figure \ref{fig:fig_nagae}, we use data from \citet{nagae04}, who presented a series of simulations for both adiabatic ($\gamma = 5/3$) and isothermal ($\gamma = 1.01$) cases with a mass ratio of $q = 0.5$. The figure compares the predictions of $\eta$ and $\eta_\mathrm{BHL}$ for a circular orbit as a function of $w$. As can be seen, incorporating the geometric factor into the standard BHL implementation significantly improves the agreement with the numerical results. Our model accurately captures the flattening of the accretion efficiency observed in the simulations for low wind-to-orbital velocity ratios. The figure also demonstrates the two asymptotic behaviors of $\eta$ established by Equations~(\ref{eq:limit1}) and (\ref{eq:limit2}), and highlights the inadequacy of the standard BHL implementation ($\eta_\mathrm{BHL} > 1$) for low $w$. Notably, good agreement persists even for points with $w<0.15$, which, according to Equation~(\ref{eq:l1}), lie outside the strict scope of our prescription.

More recently, \citet{Saladino19} presented a series of SPH simulations of accreting binaries covering a wide range of parameters (see their table 3). As expected, they found that for models with large $w$, the resulting efficiency closely matches the prediction of the standard BHL implementation ($\eta_\mathrm{sim} \approx \eta_\mathrm{BHL}$). However, for low $w$, they observed that $\eta_\mathrm{sim}$ never exceeds 30\%. To incorporate these findings into a binary population synthesis code, \citet{Saladino19} fitted their low-$w$ ($w < 1$) data with constant values.

\begin{figure}
\begin{center}
\includegraphics[width=\linewidth]{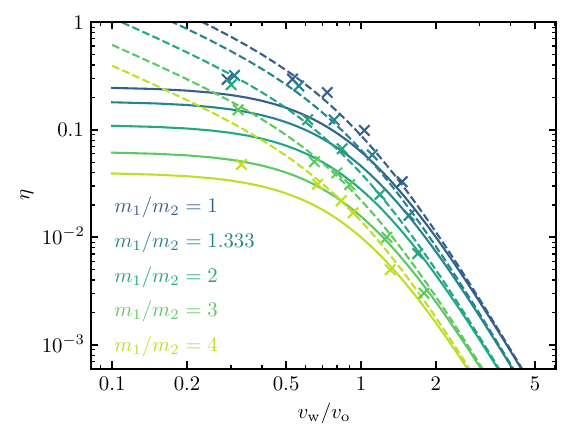}
\end{center}
\caption{Mass accretion efficiency $\eta$ predicted by our modified BHL implementation calculated for different mass ratios. The (cross) symbols represent the results obtained from the numerical simulations presented in \citet{Saladino19}. The dashed lines are the predictions of the standard BHL implementation ($\eta_\mathrm{BHL}$).}
\label{fig_saladino}
\end{figure}

Figure~\ref{fig_saladino} compares our model's predictions with the results from \citet{Saladino19}. Remarkably, our modified BHL implementation naturally predicts the flattening of the mass accretion efficiency observed in the simulations for $w < 1$, without the need for any fitting procedures. While the simulations show a similar trend to our model, slight differences are apparent, likely due to the inclusion of cooling physics in the simulations, which our model does not incorporate. A more detailed comparison with the simulations of \citet{Saladino19} is limited because they report wind velocities at the accretor's location for only one of their models (see Appendix~\ref{sec:wind} for more details). This introduces an additional source of discrepancy, as our model relies on these specific velocities.
\\
\\

\subsection{Elliptic orbits}
\label{subsec:elliptical}

For a general eccentric case, the mass accretion efficiency $\eta$ is given by \eq{eta}. We can rewrite this expression in terms of dimensionless quantities as
\begin{equation}
\begin{split}
    \eta = &q^2 \left|1-\frac{e\sin\varphi}{w(\varphi)\sqrt{1-e^2}}\right|\,\left(\frac{1+e\cos\varphi}{1-e^2}\right)^2 \times \\
    & \left[ w(\varphi)^2 +\frac{1+e^2+2\,e\cos\varphi}{1-e^2} - \frac{2 w(\varphi)\,e\sin\varphi}{\sqrt{1-e^2}} \right]^{-2},
    \label{eq:etaecc}
\end{split}    
\end{equation}
with $q$ as defined in \eq{eq:q} and $w(\varphi)$ now a function of the orbital phase that, according to \eq{eq:beta}, we can write as 
\begin{equation}
    w(\varphi) = w_\mathrm{i} + (w_\infty - w_\mathrm{i})\left[1 - \frac{R(1+e\cos\varphi)}{a(1-e^2)}\right]^\beta,
    \label{eq:wbeta}
\end{equation}
with $w_\infty = v_\infty/\vo$ and $w_\mathrm{i} = v_\mathrm{i}/\vo$.

From \eq{eq:etaecc}, we observe that $\eta$ varies periodically throughout the orbit. To characterize the overall accretion behavior, we determine the average mass accretion efficiency $\eta_\mathrm{avg}$ by integrating $\eta$ over a full orbital period $T$ as
\begin{equation}
\begin{split}
    \eta_\mathrm{avg} & = \frac{1}{T} \int_0^T \eta\, \mathrm{d} t = \frac{1}{T} \int_0^{2\pi} \eta \frac{r^2}{h} \mathrm{d}\varphi \\
    & = \frac{(1-e^2)}{2\pi}^{3/2} \int_0^{2\pi} \frac{\eta}{(1+e\cos\varphi)^2} \mathrm{d}\varphi.
    \label{eq:averaged}
\end{split}
\end{equation}

\subsubsection{Representative parameters for high-mass X-ray binaries}

\begin{figure*}
\begin{center}
\includegraphics[width=0.5\linewidth]{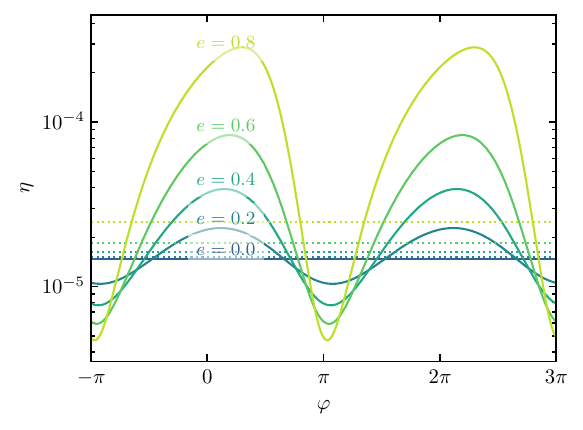}~
\includegraphics[width=0.5\linewidth]{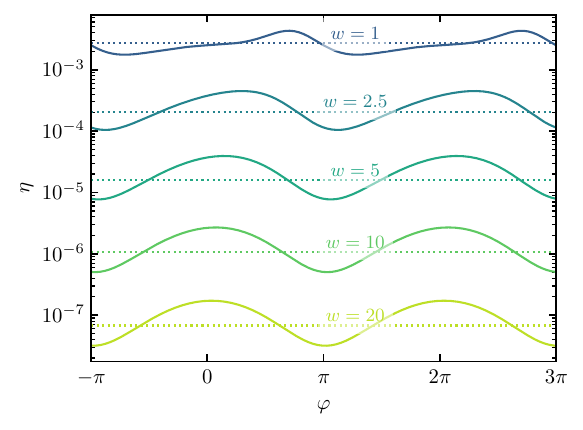}
\caption{Mass accretion efficiency $\eta$ for elliptical orbits with $q=0.1$  as a function of orbital phase. {\it Left}: Models with varying eccentricity $e$ and a fixed wind-to-orbital velocity ratio of  $w=5$. {\it Right}: Models with varying  $w$  and a fixed eccentricity of $ e=0.4$. The corresponding parameter values for each curve are labeled in the respective panels. The average mass accretion efficiency $\eta_\mathrm{avg}$ is indicated by dotted horizontal lines for each model.}
\label{fig:mdot_vel}
\end{center}
\end{figure*}

\begin{figure}
\begin{center}
\includegraphics[width=\linewidth]{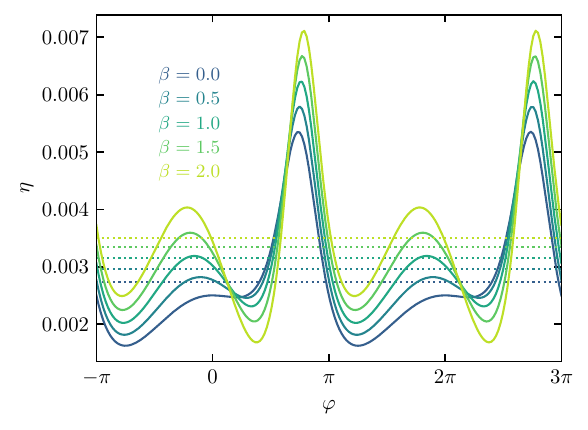}
\caption{Mass accretion efficiency $\eta$ for elliptical cases with $q = 0.1$, $e=0.5$, $v_\mathrm{i}=0$, and $w_\infty=1$, showcasing the influence of different wind acceleration profiles. The wind acceleration is modeled by \eq{eq:wbeta} with varying values of $\beta = \,0,\,0.5,\,1,\,1.5,\,2$, and a fixed wind launching radius of $R=0.1\,a$.}
\label{fig:Mdot_beta}
\end{center}
\end{figure}

Figure \ref{fig:mdot_vel} presents examples of eccentric models with a fixed mass ratio of $q = 0.1$, where we have considered that the wind has already reached its asymptotic velocity for the  spatial domain of interest (that is, $\beta=0$). The figure is divided into two panels: the left panel explores the effect of varying the  eccentricity $e$ while keeping the wind-to-orbital velocity ratio fixed $w=5$, and the right panel examines the influence of varying $w$ for a fixed eccentricity of $e = 0.4$. These values are representative of typical HMXB systems (see Figure \ref{fig:param_space} and Appendix \ref{sec:obs}).

From Figure~\ref{fig:mdot_vel}, we can observe that the resulting curves for the mass accretion efficiency are more symmetric and sinusoidal-like for cases with small $e$ (left panel) or large $w$ (right panel). In these cases, the maxima and minima of the curves correspond quite closely to the positions of periastron ($\varphi = 0$) and apastron ($\varphi = \pi$), respectively. However, as $e$ increases and $w$ decreases, the curves become increasingly asymmetric, confirming the trend observed in previous works \citep{Okuda77}. The maximum shifts notably towards higher $\varphi$ values (never exceeding $\varphi=\pi$), creating an asymmetric shape with a shallower rise towards the peak and a steeper decline after it. This effect is particularly pronounced for $e=0.8$ in the left panel of Figure~\ref{fig:mdot_vel} and the $w=1$ case of the right panel. In general, the position of the minima in the mass accretion efficiency curves presented in Figure~\ref{fig:mdot_vel} are less affected. This asymmetry arises from the complex interplay between the accretor's velocity, the wind velocity, and the varying distance between the stars along the eccentric orbit.

\begin{figure*}
\begin{center}
\includegraphics[width=0.5\linewidth]{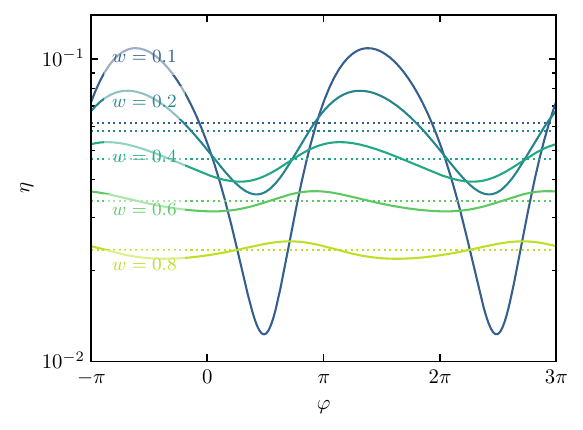}~
\includegraphics[width=0.5\linewidth]{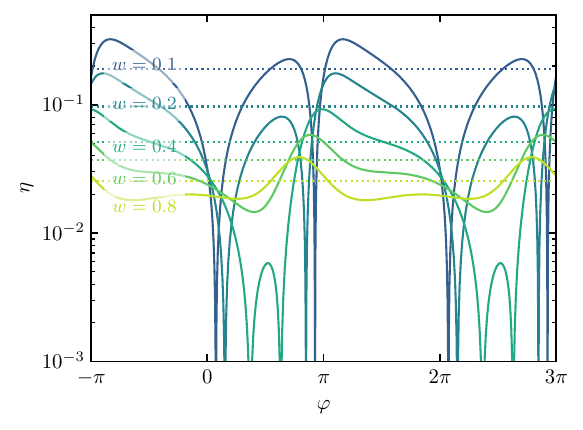}
\end{center}
\caption{Mass accretion efficiency $\eta$ for elliptical cases with $q = 0.25$ as a function of orbital phase for different $w$ values as labeled in the figure. The left and right panels show the results for $e=0.08$ and $e=0.40$, respectively. The averaged mass accretion efficiency ($\eta_\mathrm{avg}$) is shown with dotted lines for all models.}
\label{fig:mdot_ss}
\end{figure*}

Focusing now on the left panel of Figure~\ref{fig:mdot_vel}, we see that, as the eccentricity increases, the peak-to-valley ratio of the curves also increases, spanning almost two orders of magnitude for the $e = 0.8$ model. The averaged mass accretion efficiency $\eta_\mathrm{avg}$, illustrated with dotted horizontal lines, also increases with $e$. This trend arises because, even though the relative velocity is larger close to periastron and the accretor spends less time there, the mass accretion is more efficient due to the enhanced wind density in this region. This enhanced density outweighs the effects of increased velocity and reduced time of passage, leading to a net increase in the average accretion efficiency with eccentricity.

The right panel of Figure~\ref{fig:mdot_vel} shows the variations in mass accretion efficiency for different values of $w$, with the eccentricity fixed at  $e=0.4$. In agreement with \eq{eq:limit1}, as $w$ increases, the mass accretion becomes less efficient. For instance, a model with $w=20$ has an average mass accretion efficiency of  $\eta_\mathrm{avg}\approx7\times10^{-8}$, while a model with  $w=2.5$  results in  $\eta_\mathrm{avg}\approx3\times10^{-4}$. 

In Figure~\ref{fig:Mdot_beta}, we delve into the impact of different wind acceleration profiles on $\eta$. To isolate this effect, we fix the mass ratio $q$, eccentricity $e$, initially wind velocity $v_\mathrm{i}$, and wind-to-orbital velocity ratio $w_\infty = v_\infty/\vo$ to representative values, and explore the influence of the wind acceleration parameter $\beta$ which ranges from 0 (instantaneous acceleration) to 2 (gradual acceleration). As the figure demonstrates, while the specific value of  $\beta$  does not significantly affect the average mass accretion efficiency, it can substantially alter the shape of the efficiency curve.  A smoother wind acceleration ($\beta$  closer to 2) leads to the emergence of two additional extremal points: a local maximum and a local minimum.  Moreover, as  $\beta$  increases, the ratio between the global maximum and minimum also increases, and both extrema appear to shift closer to apastron, resulting in an abrupt rise and drop in the mass accretion efficiency at that orbital position.

\subsubsection{Representative parameters for symbiotic systems}

Figure~\ref{fig:mdot_ss} presents calculations for elliptical models with $q=0.25$, $\beta = 0$, and varying $w$ values, chosen to be representative of typical symbiotic systems (see Figure~\ref{fig:param_space} and Appendix~\ref{sec:obs}). The left panel focuses on models with an eccentricity of $e=0.08$, which is the median value obtained from all confirmed galactic symbiotic systems in the {\it New Online Database of Symbiotic Variables} with available information \citep{Merc2019}. The right panel, in contrast, explores a higher eccentricity of $ e=0.4 $ to illustrate the impact of more eccentric orbits on the accretion process.

As seen in the left panel, the average mass accretion efficiency $\eta_\mathrm{avg}$ increases as $w$ decreases, as predicted by the BHL model. Interestingly, even with the relatively modest eccentricity of $e=0.08$, the range of mass accretion efficiency values increases significantly as  $w$ decreases, as evidenced by the growing ratio between the maximum and minimum values. This highlights the impact of even small eccentricities on the accretion process, particularly for slower winds.

Furthermore, the figure reveals that both the minimum and maximum efficiency shift to smaller orbital phases $\varphi$ as $w$ increases. For instance, with $w=0.1$, the maximum is reached after apastron at $\varphi \simeq 4.4$, whereas for  $w=0.8$, it occurs before apastron at $\varphi \simeq 2.2$.

\begin{figure*}
\begin{center}
\includegraphics[width=0.5\linewidth]{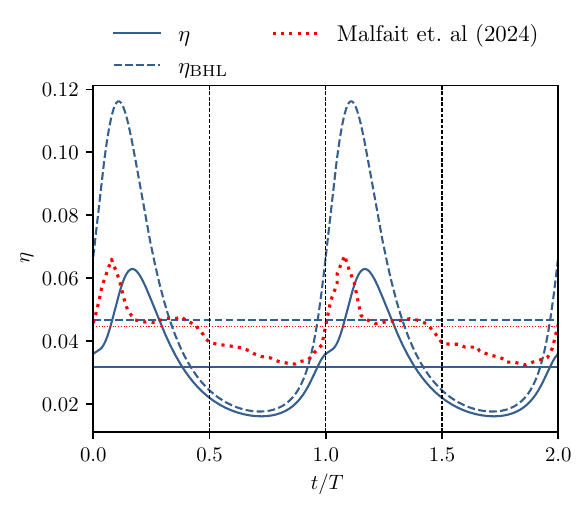}~
\includegraphics[width=0.5\linewidth]{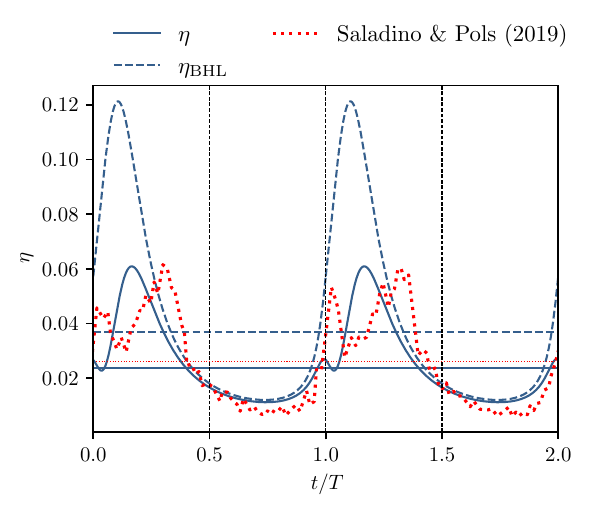}
\caption{Comparison of the mass accretion efficiencies $\eta$ (solid) and $\eta_\mathrm{BHL}$ (dashed) with estimations from numerical simulations $\eta_\mathrm{sim}$ (red dotted) for eccentric orbits. The horizontal axis shows time $t$ in units of the orbital period $T$. Vertical dashed lines indicate the location of periastron ($t=0$) and apoastron ($t=0.5 T$). Left: A model with $e=0.5$, $v_\mathrm{i}=20$ km~s$^{-1}$, $v_\infty=24.8$ km~s$^{-1}$, and $\beta=2.65$ from \citet{Malfait2024}. Right: A model with $e=0.6$, $v_\mathrm{i}=12$ km~s$^{-1}$, $v_\infty=15.4$ km~s$^{-1}$, and $\beta=0.88$ from \citet{SaladinoPols2019}. See further details in Table~\ref{tab:comparison} and Appendix~\ref{sec:wind}.}
\label{fig:malfail}
\end{center}
\end{figure*}

On the right panel of Figure~\ref{fig:mdot_ss}, where we consider a higher eccentricity of  $e=0.4$, we observe similar trends to those in the left panel. However, a new qualitative feature emerges: for sufficiently small values of $w$, the mass accretion efficiency completely vanishes at two points in the orbital phase. This intriguing behavior arises from the geometric projection factor in \eq{eta}, which becomes zero whenever the wind velocity ($ \vw$) equals the radial component of the accretor's velocity ($\va^r$). At these specific points, the wind direction is instantaneously perpendicular to the accretion cylinder, resulting in no material being captured. Note that this condition does not imply that the accretor is co-moving with the wind, but only that its radial velocity component is instantaneously balanced by the local wind velocity, leaving the accretor with a purely tangential velocity still within the supersonic regime of the BHL framework.

Using \eq{eq:etaecc}, the general condition for the vanishing of mass accretion efficiency can be expressed as 
\begin{equation}
   \frac{e\sin\varphi}{\sqrt{1-e^2}} = \frac{v_\mathrm{i}}{\vo} + \left(\frac{v_\infty - v_\mathrm{i}}{\vo}\right)\left[1 + \frac{R(1+e\cos\varphi)}{a(1-e^2)}\right]^\beta. 
 \end{equation}
For an arbitrary value of $\beta$, this equation requires numerical methods to find solutions. However, we can derive analytical solutions for the specific case where the wind has already reached its terminal velocity $\vw = v_\infty$  throughout the relevant domain ($\beta = 0$). In this scenario, the mass accretion efficiency vanishes if
\begin{equation}
    w \leq \frac{e}{\sqrt{1-e^2}},
\end{equation}
\noindent which occurs at the following two orbital phases
\begin{gather}
    \varphi_1 = \arcsin(w\sqrt{1-e^2}/e), \label{eq:phi1}\\
    \varphi_2 = \pi - \varphi_1.
    \label{eq:phi2}
\end{gather}
These two points first appear at $\varphi_1=\varphi_2=\pi/2$ when $w = e/\sqrt{1-e^2}$. As  $w$  decreases below this threshold ($w < e/\sqrt{1-e^2}$), the points split symmetrically, moving away from $\pi/2$ and always satisfying $0<\varphi_1 <\varphi_2 <\pi$, as seen in the right panel of Figure~\ref{fig:mdot_ss}). This behavior indicates that these points of vanishing accretion efficiency emerge exclusively during the accretor's motion from periastron to apastron.

While a complete cancellation of the accretion efficiency is unlikely under realistic astrophysical conditions with clumps, shocks, and velocity variations in stellar winds, our model suggests that, under appropriate circumstances with sufficiently high eccentricities and low wind velocities, one could expect orbital phases with a sharp decline in accretion. Provided these drops are reflected in the resulting emission, this potentially testable prediction offers a unique observational signature of our modified BHL implementation.

\subsubsection{Impact of a detailed wind profile}

Determining the wind acceleration profile in a binary is difficult due to complex dynamical structures that form around the accretor such as shocks, accretion disks, and spiral patterns (see the numerical works listed in Table~\ref{tab:comparison}). To circumvent this, we use the unperturbed wind profile of the donor as a reasonable first approximation. We validate this by comparing our model's predicted accretion efficiencies with those from simulations of eccentric orbits. For these comparisons, we use wind profiles from simulations of single mass-losing stars, where complex structures are absent. Details of our profile modeling are in Appendix~\ref{sec:wind}.

Figure~\ref{fig:malfail} shows the evolution of $\eta$ for eccentric orbits with parameters similar to those in \citet{Malfait2024} and \citet{SaladinoPols2019}. Our model captures the general trend of the accretion history, though fine details differ. This likely arises from complexities not included in our model, such as wind structure variations and mass reprocessing within the accretion disk, as seen in the simulations (see Section~\ref{sec:caveats}).

Finally, we note that discrepancies are most pronounced when the accreting companion lies within the wind acceleration zone of the donor star. This occurs at low wind velocities in \citet{Malfait2024} and small eccentricities in \citet{SaladinoPols2019} (see Figure~\ref{fig:wind_malfait}).

\section{Discussion}
\label{sec:discussion}

Understanding the accretion process is fundamental to advancing our knowledge of the evolution of interacting binaries, as it significantly influences the evolutionary path of both stars. Accretion also directly affects the observable properties of these systems, such as the luminosity of symbiotic stars and HMXBs, which can be directly compared with optical and X-ray observations \citep[see for example][]{Pujol2023,Toala2024,Zamanov2024}.

However, as discussed in Section~\ref{sec:intro}, the standard implementation of the BHL model faces limitations when the wind velocity is less than the orbital velocity ($w<1$), a regime relevant to most symbiotic systems. In this regime, the standard BHL implementation (Equation~\ref{etaBHL}) can predict non-physical accretion efficiencies exceeding unity, an inadequacy corroborated by numerical studies \citep[e.g.,][]{nagae04,Saladino19,ElMellah19}. This has lead to the prevailing view that the BHL model is only applicable  in the high wind-to-orbital velocity regime  \citep[$w \gg 1$; see, e.g.,][]{Boffin15,Hansen2016}.

In this work, we have addressed this limitation by incorporating a geometric projection factor within the standard BHL framework, extending its applicability to the regime $w\leq1$. This modification, derived from simple geometric considerations, provides a unified and physically motivated description of wind accretion in binary systems, accurately capturing the accretion dynamics across a wide range of wind-to-orbital velocity ratios.

For circular orbits, our modified model (Equation~\ref{eta_c}) predicts that the accretion efficiency $\eta$ depends solely on the dimensionless ratios $q$ and $w$, always guaranteeing physically plausible values $\eta \leq 1$. It recovers the scaling of the standard BHL implementation $\eta=q^2/w^4$ for $w\gg1$ (Equation~\ref{eq:limit1}) and predicts a constant efficiency $\eta = q^2$ for $w\ll1$ (Equation~\ref{eq:limit2}), consistent with numerical simulations.

This improved accuracy has significant implications for modeling binary systems.  Previous studies, such as those by \citet{Vos2015}, \citet{Abate2015}, and \citet{Yungelson19}, relied on accretion prescriptions tailored to the  $w\gg1$  regime. By incorporating our modified BHL formalism, which accurately captures the accretion efficiency across the full range of  $w$, future models will be able to achieve more precise predictions of binary system evolution, including the circularization process of elliptical orbits \citep[e.g.,][]{Bonacic08,SaladinoPols2019,Misra23}. Furthermore, our approach offers a simplification compared to the more intricate prescriptions of \citet{Saladino19}, while maintaining broad applicability.

\subsection{Applications: two representative examples}

To illustrate the versatility of our modified implementation of the BHL model, we now apply it to two distinct types of accreting binary systems: symbiotic stars and HMXBs. As shown in Figure \ref{fig:param_space}, these systems occupy different regions of the parameter space. HMXBs typically reside in the high wind-to-orbital velocity ratio regime ($w>1$), where the classic BHL approximation remains valid. In contrast, symbiotic stars are generally found in the  $w\lesssim1$ regime, where the standard BHL implementation overestimates the accretion efficiency and the geometric correction becomes crucial. Furthermore, symbiotic systems exhibit higher mass accretion efficiencies ($\eta=0.01-0.1$) compared to HMXBs ($\eta=10^{-8}$--$10^{-4}$), indicating a more efficient capture of material from their mass-losing companions.

\subsubsection{R Aquarii: a symbiotic system}

\begin{figure}
\begin{center}
\includegraphics[width=\linewidth]{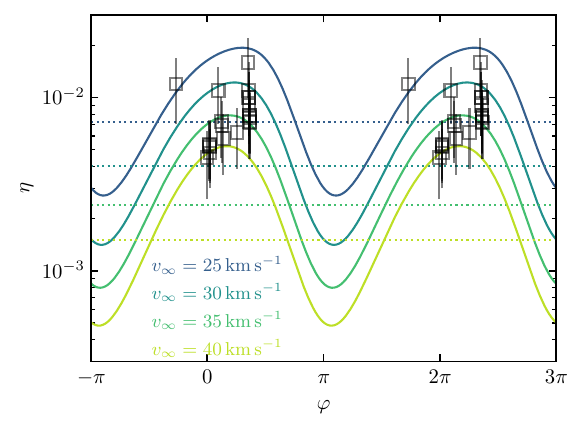}
\end{center}
\caption{Prediction of the mass accretion efficiency $\eta$ for the symbiotic system R~Aqr. Results are shown for different terminal wind velocities. The dotted lines represent the $\eta_\mathrm{avg}$ values obtained for the complete orbit. The squares (and error bars) are estimations obtained from optical observations analyzed in \citet{Vasquez-Torres24} around periastron passage.}
\label{fig:RAq}
\end{figure}

Figure \ref{fig:RAq} presents our model's predictions for the evolution of the mass accretion efficiency in the symbiotic system \mbox{R Aqr}, one of the best-characterized systems of this type. \mbox{R Aqr} has an orbital period of 42 years, with stellar component masses of $m_1=1.0\Msun$ and $m_2=0.7\Msun$, and an eccentricity of $e=0.45$ \citep[see][and references therein]{Alcolea2023}. To account for the variable nature of the wind in this late-type star, we adopted a mass-loss rate of  $\dot{M}_\mathrm{w}=10^{-7}$ M$_\odot$~yr$^{-1}$ \citep{Michalitsianos1980,Spergel1983} and considered wind velocities of 25, 30, 35, and 40 km s$^{-1}$. Our model predicts corresponding average mass accretion efficiencies of $\eta_\mathrm{avg}=$\num{7.2e-3}, \num{4.0e-3}, \num{2.4e-3}, and \num{1.5e-3}, respectively. 
Furthermore, we estimate that the mass accretion efficiencies during periastron for these parameters are between $4.7\times10^{-3}$ and $1.6\times10^{-2}$, which agree with predictions from the analysis of observations of R Aqr recently presented in \citet{Vasquez-Torres24}.

The simulations presented by \citet{Vathachira2024} of an evolving 1~M$_\odot$ star with an accreting companion with initial mass of 0.7 M$_\odot$ at an orbital separation of 15.9~au are very similar to the configuration of R Aqr. These authors predict a more or less constant mass accretion efficiency of \mbox{$\approx 6.5\times10^{-3}$} \citep[see figure 3 in][]{Vathachira2024}, consistent with our averaged predicted value for R Aqr with a wind of 25 km~s$^{-1}$. However, the calculations presented by those authors can not reproduce the expected variable accretion history of R Aqr, or the expected accretion efficiency during periastron passage, given that their simulations are tailored for circular orbits.

\begin{figure}
\begin{center}
\includegraphics[width=\linewidth]{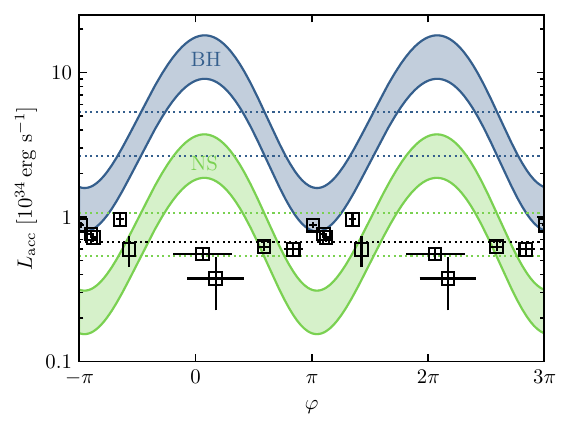}
\end{center}
\caption{Comparison of the accretion luminosity predicted by our model for a black hole (BH; blue region) and a neutron star (NS; green region) accretor in LS~5039 with observational data from \citet{Zabalza2008}. The wind parameters are $v_\infty = 2440$~km~s$^{-1}$ and $\beta= 0.8$, and the mass-loss rate ranges from $\dot M_\mathrm{w} = 1 \times 10^{-7}$ to $2 \times 10^{-7} \Msun$~yr$^{-1}$. The lower and upper bounds of each shaded region correspond to these limiting values of $\dot{M}_\mathrm{w}$. The average luminosities for the BH and NS scenarios are $2.66 \times 10^{34}$ and $5.36 \times 10^{33}$~erg~s$^{-1}$, respectively.  The observed average luminosity is $6.75 \times 10^{33}$~erg~s$^{-1}$.}
\label{fig:HMXRB}
\end{figure}

\subsubsection{LS 5039: a high-mass X-ray binary}

We now turn our attention to LS~5039, a HMXB system whose compact accretor has been the subject of much debate \citep{Casares2005,Yoneda2020,Volkov2021,Makishima2023}. Multi-epoch X-ray observations have revealed an average flux of $F_\mathrm{X}\lesssim10^{-11}$~erg~cm$^{-2}$~s$^{-1}$ in the 1--10 keV band \citep{Zabalza2008,Takahashi2009}, which, at a distance of 2.5 kpc, translates to an average X-ray luminosity of $L_\mathrm{X,avg}\approx7\times10^{33}$~erg~s$^{-1}$. 

Detailed orbital analysis of LS 5039 suggests an eccentricity of $e=0.35$, a period of $T=3.9$ days, and a massive O-type companion star with a mass of $m_1=22.9\Msun$ and a radius of 9.3 R$_\odot$ \citep{Casares2005}. The nature of the compact object remains uncertain, with possibilities including a black hole (BH) of mass $m_\mathrm{BH}=3.7\Msun$  or a neutron star (NS) of mass $m_\mathrm{NS}=1.4\Msun$  and radius $R_\mathrm{NS}=10$ km \citep{McSwain2002,Casares2005}.  

To investigate the nature of the accretor, we apply our modified BHL model, adopting a primary's wind with terminal velocity of 2440~km~s$^{-1}$, a mass loss rate $ \dot M_\mathrm{w}$ of 1--2$\times$10$^{-7}$, and $\beta=0.8$ \citep{Casares2005,Dubus2015,Reig03,Kissmann2023}. We estimate the resulting accretion luminosity $L_\mathrm{X}$ assuming that the accreted material forms a thin accretion disk around the compact object. We use the standard expression for this process \citep{SS73}:  
\begin{equation}
L_\mathrm{X} = \frac{1}{2}\frac{Gm_2}{r_\mathrm{in}} \dot M_\mathrm{acc}, 
\end{equation}
where $\dot M_\mathrm{acc} = \eta \dot M_\mathrm{w}$, $m_2$ is the mass of the accreting object, and $r_\mathrm{in}$ is the inner radius of the accretion disk. For a weakly magnetized NS, we can take $r_\mathrm{in} = R_\mathrm{NS} = 10\,$km, while for a non-rotating BH, we take the radius of the innermost stable circular orbit: $r_\mathrm{in} = 6 G m_\mathrm{BH}/c^2 \simeq 32.9\,$km.

Figure \ref{fig:HMXRB} compares the accretion luminosity predicted by our model for both a black hole (blue region) and a neutron star (green region) accretor with observational data from \citet{Zabalza2008}. The lower and upper bounds of each region correspond to mass loss rates of $10^{-7}$  and  $2\times10^{-7}$~M$_\odot$~yr$^{-1}$, respectively. The figure shows that the average luminosity for a neutron star accretor \mbox{($5.4\times10^{33}$erg\,s$^{-1}$)} agrees better with the averaged observed value than that for a black hole accretor \mbox{($2.7\times10^{34}$erg\,s$^{-1}$)}.

However, our model predicts a wider range in luminosity ($L_{\max}/L_{\min}=8.3$) than inferred from observations ($L_{\max}/L_{\min}=2.6$). Moreover, our model predicts a maximum luminosity around periastron and a minimum around apastron, while the observations suggest the opposite trend. This discrepancy may stem from the fact that our model does not include a detailed treatment of the accretion disk. The reprocessing of accreted material within the disk significantly influences the conversion of mass accretion into luminosity, potentially altering the phase dependence and the range of the observed luminosity (see next subsection).

These findings lend support to the hypothesis that the compact object in LS 5039 is indeed a NS, consistent with previous studies that have suggested a similar conclusion based on X-ray observations \citep[e.g.,][]{Makishima2023}. However, further investigations, incorporating a more detailed treatment of the accretion disk physics and radiative processes, are needed to definitively determine the nature of the compact object in LS 5039.

\subsection{Caveats}
\label{sec:caveats}

While our modified prescription of the BHL model offers a more accurate and comprehensive description of wind accretion in binary systems than its standard implementation, it's essential to acknowledge certain limitations and outline future refinements. In this section, we discuss these caveats, highlight potential avenues for further development, and address discrepancies between our model and existing numerical simulations.

\subsubsection{Vanishing mass accretion rate}

One key prediction of our model is the complete vanishing of mass accretion efficiency at specific orbital phases for eccentric orbits with sufficiently low wind-to-orbital velocity ratios. However, this prediction relies on an idealized scenario of a smooth and homogeneous wind. In reality, stellar winds are variable and exhibit density and velocity inhomogeneities, which would prevent such a complete cancellation of accretion. Nonetheless, our model suggests that even with these inhomogeneities, systems with high eccentricities and low wind velocities could still exhibit sharp drops in accretion at certain orbital phases. Whether these drops translate into observable features requires further investigation, considering the specific radiative processes and the interaction of the emitted radiation with the circumbinary environment.

Despite these caveats, preliminary analysis of optical spectra from the symbiotic system BX Mon ($e=0.45$) suggests that its variability can be explained by curves similar to those presented in Figure \ref{fig:mdot_ss} (Toalá et al., in prep.), providing tentative observational support for our prediction of sharp variations in accretion efficiency for eccentric orbits.

\subsubsection{Comparison with simulations}

Our model generally shows good agreement with numerical simulations (Table \ref{tab:comparison}). However, some discrepancies arise, and we identify three potential sources:

\begin{enumerate}
    \item {\bf Wind acceleration:} We find better agreement with simulations where the wind reaches its terminal velocity before encountering the accretor \citep[e.g.,][]{Theuns96,ValBorro09,HE13,Chen2020}, cases where the wind acceleration is almost instantaneous \citep[][]{Lee2022,Malfait2024} and those with high eccentricity \citep{SaladinoPols2019}. Discrepancies occur when the wind is still accelerating, making it difficult to define a precise wind velocity for comparison. This sensitivity to the wind velocity and acceleration profile, consistent with our findings in Figures~\ref{fig:mdot_vel} and \ref{fig:Mdot_beta}, highlights the importance of carefully considering the wind properties when applying our model, especially in scenarios with poorly constrained or highly variable winds.

\item {\bf Accretion disks:} Our modified BHL implementation does not include accretion disks, but several of the simulations listed in Table \ref{tab:comparison} do, with accretion rates calculated at a radius much smaller than the characteristic BHL accretion radius $r_\mathrm{acc}$ (Equation~\ref{racc}). This difference in treatment can lead to discrepancies, as the disk can reprocess the accreted material, potentially modifying the accretion history \citep[see e.g.,][]{Chen17}. Figure \ref{fig:malfail} compares our model's predictions with simulations of eccentric cases from \citet{Malfait2024} and \citet{SaladinoPols2019}, showing that while our model captures the average accretion efficiency well (see also Table~\ref{tab:comparison}), the detailed accretion histories differ. To facilitate more direct comparisons, future simulations could track the accretion history at a radius comparable to $r_\mathrm{acc}$. This would provide a more straightforward way to assess our model's accuracy in capturing the time-dependent behavior of the accretion process.

\item {\bf  Inherent limitations and numerical effects:} Numerical simulations are subject to limitations and uncertainties related to numerical resolution, algorithms, and boundary conditions, which could contribute to the observed discrepancies. However, it is also important to note that the level of agreement between our model and simulations (typically within a factor of a few) is similar to that observed for the classic BHL model within its regime of applicability \citep{hunt71,shima85,ruffert94}, where a dimensionless correction factor ($\alpha\lesssim 1$) is often introduced to reconcile the model with simulations. This disparity is attributed to various factors, including accretor's size, limitations in numerical resolution, the equation of state of the accreting gas, and the inclusion of more complex physical processes.

\end{enumerate}

By incorporating a geometric correction factor, our modified implementation of the BHL model provides an accurate description of mass transfer rates in wind-fed binary systems, especially for low wind-to-orbital velocity ratios. However, some discrepancies remain between our predicted accretion rate histories and those found in numerical simulations. These discrepancies highlight the complex nature of wind accretion and emphasize the need for further investigations to improve both theoretical models and numerical simulations, ultimately leading to a more comprehensive understanding of wind accretion processes in binary systems. 

This work establishes a baseline for exploring more complex accretion physics in binary systems, such as the impact of inhomogeneities in the wind \citep{Ducci09}, disk formation \citep{HE13,Chen17}, colliding-wind binaries \citep{Kashi2022}, and the primary's periodic cutting of the accretion wake \citep{Kashi2023}. Finally, improving our understanding of the accretion process is crucial to understand the formation of Barium stars \citep{Bidelman1951} and carbon- and $s$-element-enhanced metal-poor stars \citep{Beers2005}.

\section{Summary}
\label{sec:summary}

We revisited the implementation of the BHL accretion model for binary systems, focusing on the regime where the wind velocity $v_\mathrm{w}$ is comparable to or less than the orbital velocity $v_\mathrm{o}$. By introducing a geometric correction factor, we derived a modified expression for the mass accretion efficiency $\eta$ that remains physically plausible ($\eta \leq 1$) even for $w =\vw/\vo \ll 1$, addressing a key limitation of the standard implementation of the BHL model.

Our modified model predicts two distinct asymptotic behaviors for circular orbits:  $\eta= q^2/w^4$  for  $w\gg1$, recovering the well-known scaling of the standard BHL implementation, and  $\eta= q^2$  for  $w\ll1$, consistent with numerical simulations. This accurate description across the full range of  $w$, without the need for {\it ad hoc} adjustments, highlights the strength of our geometrically corrected BHL implementation.

We applied our model to both symbiotic systems and HMXBs, finding that symbiotic systems exhibit higher accretion efficiencies than HMXBs (Figure~\ref{fig:param_space}). Our analysis of eccentric orbits revealed a complex interplay between eccentricity $e$, the wind-to-orbital velocity ratio  $w$, and the wind acceleration parameter $\beta$, leading to significant variations in  $\eta$. Notably, we predict the possibility of sharp drops in accretion at specific orbital phases for sufficiently eccentric systems with low  $w$, offering a potential observational signature of our model.

Our modified BHL implementation provides a more accurate and comprehensive framework for understanding wind accretion in binary systems, with implications for stellar evolution, population synthesis, and the interpretation of observational data. 
\\


\noindent The authors thank comments and suggestions by an anonymous referee that improved the clarity of the original manuscript. They also thank Jolien Malfait for providing 1D wind velocity profiles. J.A.T.~acknowledges support from the UNAM PAPIIT project IN102324. This work has made extensive use of NASA's Astrophysics Data System (ADS). 


\bibliography{references}{}
\bibliographystyle{aasjournal}

\appendix

\section{Observed properties of binary systems}
\restartappendixnumbering
\label{sec:obs}

Table~\ref{tab:objects} provides key observational properties of selected symbiotic and high-mass X-ray binary (HMXB) systems used to exemplify our theoretical model predictions. These systems, characterized  by distinct wind velocities and mass accretion properties, are also plotted in the parameter space shown in Figure~\ref{fig:param_space}. The data for the symbiotic systems were sourced from the {\it New Online Database of Symbiotic Variables} \citep{Merc2019}\footnote{\url{https://sirrah.troja.mff.cuni.cz/~merc/nodsv/}}, while HMXB properties were taken from {\it A Catalogue of High-Mass X-ray Binaries in the Galaxy From the INTEGRAL to the Gaia era} \citep{Fortin2023}\footnote{\url{https://binary-revolution.github.io/HMXBwebcat/}}. 

Table~\ref{tab:objects} also includes the predicted mass accretion efficiencies $\eta$ and $\etaBHL$ computed for each system which, for simplicity, have been computed assuming circular orbits. For symbiotic systems, we assume a typical stellar wind velocity of 20~km~s$^{-1}$, with a possible range from 10 to 30~km~s$^{-1}$ \citep{Ramstedt2020}. For OB stars in HMXB systems, a standard velocity of 2000~km~s$^{-1}$ is adopted, with a range of 1000--3000~km~s$^{-1}$ \citep{Vink2021,Hawcroft2024}.

Notably, with its extended orbital period, $o$ Ceti is positioned closer to the HMXB region in Figure~\ref{fig:param_space} due to its relatively low orbital velocity ($\vo\simeq5$ km\,s$^{-1}$). The estimated mass accretion efficiency ($\eta=2.5\times10^{-4}$) implies a corresponding mass accretion rate of $\dot{M}_\mathrm{acc}=\eta \dot{M}_\mathrm{w} \approx 8\times10^{-11}$~M$_\odot$~yr$^{-1}$, assuming a mass-loss rate of $\dot{M}_\mathrm{w}=3\times10^{-7}$~M$_\odot$~yr$^{-1}$ from the late-type star, as reported in \citet{Ryde2000}. This value of $\dot{M}_\mathrm{acc}$ is consistent with estimates of $\lesssim10^{-10}$~M$_\odot$~yr$^{-1}$ from \citet{Sokoloski2010}.

\begin{table*}
\begin{center}
\footnotesize
\caption{Observed properties and estimated mass accretion efficiencies of selected symbiotic and HMXB systems. The columns list, from left to right: mass of the primary (donor), mass of the secondary (accretor), orbital period, mass ratio $q=m_2/(m_1+m_2)$, range of wind-to-orbital velocity ratio $w = \vw/\vo$, and calculated mass accretion efficiencies ($\eta$ and $\etaBHL$). For symbiotic systems, we assume a typical wind velocity of 20~km~s$^{-1}$, while for HMXBs, we take 2000~km~s$^{-1}$. These parameters inform the computed $\eta$ values (Equation~\ref{eta_c}), which are then compared with theoretical predictions from the standard BHL implementation $\etaBHL$.} 
\setlength{\tabcolsep}{0.8\tabcolsep}  
\begin{tabular}{lccccccc}
\hline
Object & $m_1$ & $m_2$ & $T$ & $q$ & $w$ & $\eta$ & $\etaBHL$ \\
  & [M$_\odot$] & [M$_\odot$] & [yr] &  &  \\
\hline
{\bf Symbiotic stars}\\
$o$ Cet &  2.00 &  0.65 & 497.90 &  0.25 &  3.84 & \num{2.42e-04} & \num{2.50e-04} \\
AE Ara &  2.00 &  0.51 &  2.00 &  0.20 &  0.62 & \num{2.15e-02} & \num{4.06e-02} \\
AG Dra &  1.20 &  0.50 &  1.51 &  0.29 &  0.65 & \num{4.31e-02} & \num{7.95e-02} \\
AG Peg &  2.60 &  0.65 &  2.23 &  0.20 &  0.59 & \num{2.19e-02} & \num{4.30e-02} \\
AR Pav &  2.00 &  0.75 &  1.65 &  0.27 &  0.57 & \num{4.27e-02} & \num{8.66e-02} \\
AX Per &  3.00 &  0.60 &  1.86 &  0.17 &  0.54 & \num{1.67e-02} & \num{3.52e-02} \\
BF Cyg &  1.50 &  0.40 &  2.07 &  0.21 &  0.69 & \num{2.03e-02} & \num{3.57e-02} \\
BX Mon &  3.70 &  0.55 &  3.78 &  0.13 &  0.65 & \num{8.34e-03} & \num{1.54e-02} \\
CH Cyg &  2.20 &  0.56 & 15.58 &  0.20 &  1.20 & \num{6.98e-03} & \num{9.10e-03} \\
CQ Dra &  5.00 &  0.85 &  4.66 &  0.15 &  0.62 & \num{1.10e-02} & \num{2.08e-02} \\
EG And &  1.46 &  0.40 &  1.32 &  0.22 &  0.60 & \num{2.51e-02} & \num{4.88e-02} \\
ER Del &  3.00 &  0.70 &  5.72 &  0.19 &  0.78 & \num{1.39e-02} & \num{2.27e-02} \\
FG Ser &  1.70 &  0.60 &  1.78 &  0.26 &  0.62 & \num{3.57e-02} & \num{6.81e-02} \\
KX TrA &  1.00 &  0.41 &  3.31 &  0.29 &  0.89 & \num{2.62e-02} & \num{3.94e-02} \\
PU Vul &  1.00 &  0.50 & 13.42 &  0.33 &  1.39 & \num{1.28e-02} & \num{1.58e-02} \\
R Aqr &  1.00 &  0.70 & 42.40 &  0.41 &  1.96 & \num{7.22e-03} & \num{8.10e-03} \\
RW Hya &  1.60 &  0.48 &  1.01 &  0.23 &  0.53 & \num{3.26e-02} & \num{6.98e-02} \\
SY Mus &  1.30 &  0.43 &  1.71 &  0.25 &  0.67 & \num{2.95e-02} & \num{5.31e-02} \\
TX CVn &  3.50 &  0.40 &  0.55 &  0.10 &  0.35 & \num{8.36e-03} & \num{2.53e-02} \\
V443 Her  &  2.50 &  0.42 &  1.64 &  0.14 &  0.55 & \num{1.21e-02} & \num{2.50e-02} \\
V694 Mon  &  1.00 &  0.90 &  5.28 &  0.47 &  0.94 & \num{6.28e-02} & \num{9.15e-02} \\
V1261 Ori &  1.65 &  0.55 &  1.75 &  0.25 &  0.62 & \num{3.25e-02} & \num{6.15e-02} \\
Z And  &  2.00 &  0.65 &  2.08 &  0.25 &  0.62 & \num{3.14e-02} & \num{5.97e-02} \\
\hline
{\bf HMXB }\\
1E 1145.1-6141 & 14.00 &  1.70 & 0.039 &  0.11 &  9.12 & \num{1.66e-06} & \num{1.67e-06} \\
1FGL J1018.6-5856 & 22.90 &  2.00 & 0.045 &  0.08 &  8.19 & \num{1.39e-06} & \num{1.40e-06} \\
4U 1538-522 & 20.00 &  1.18 & 0.010 &  0.06 &  5.26 & \num{3.78e-06} & \num{3.85e-06} \\
4U 1700-377 & 46.00 &  1.96 & 0.009 &  0.04 &  3.89 & \num{6.42e-06} & \num{6.63e-06} \\
Cen X-3 & 20.20 &  1.34 & 0.006 &  0.06 &  4.28 & \num{1.04e-05} & \num{1.07e-05} \\
Cyg X-1 & 40.60 & 21.20 & 0.015 &  0.34 &  4.22 & \num{3.34e-04} & \num{3.43e-04} \\
EXO 1722-363 & 18.00 &  1.91 & 0.027 &  0.10 &  7.40 & \num{2.96e-06} & \num{2.99e-06} \\
HD 96670 & 22.70 &  6.20 & 0.014 &  0.21 &  5.33 & \num{5.33e-05} & \num{5.42e-05} \\
HD 259440 & 15.70 &  1.40 & 0.869 &  0.08 & 24.86 & \num{1.75e-08} & \num{1.75e-08} \\
IGR J17544-2619 & 23.00 &  1.40 & 0.013 &  0.06 &  5.51 & \num{3.36e-06} & \num{3.41e-06} \\
IGR J18027-2016 & 20.00 &  1.50 & 0.012 &  0.07 &  5.60 & \num{4.65e-06} & \num{4.72e-06} \\
MWC 656 &  7.80 &  5.40 & 0.165 &  0.41 & 15.59 & \num{2.81e-06} & \num{2.82e-06} \\
OAO 1657-415 & 14.30 &  1.42 & 0.029 &  0.09 &  8.19 & \num{1.76e-06} & \num{1.77e-06} \\
SGR 0755-2933 & 18.50 &  1.40 & 0.163 &  0.07 & 13.54 & \num{1.46e-07} & \num{1.46e-07} \\
SS 433 & 11.30 &  4.20 & 0.036 &  0.27 &  8.87 & \num{1.15e-05} & \num{1.16e-05} \\
Vela X-1 & 26.00 &  2.12 & 0.025 &  0.08 &  6.41 & \num{3.20e-06} & \num{3.24e-06} \\
\hline
\end{tabular}
\label{tab:objects}
\end{center}
\end{table*}

\section{Wind acceleration}
\restartappendixnumbering
\label{sec:wind}

\begin{figure}
\begin{center}
\includegraphics[width=\linewidth]{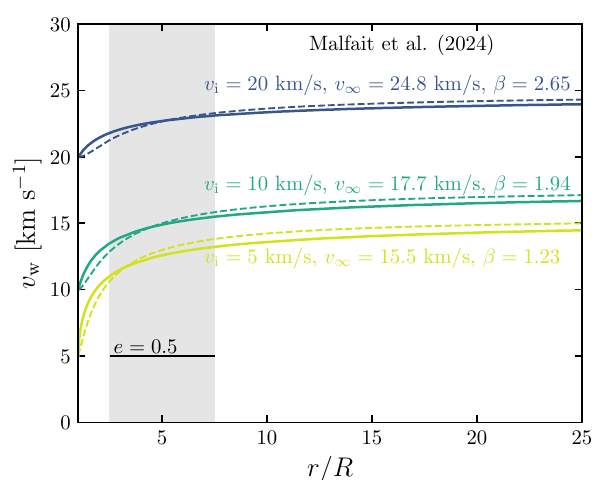}\\
\includegraphics[width=\linewidth]{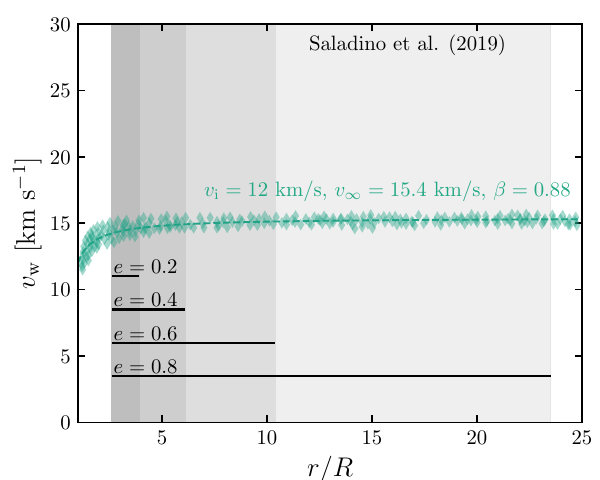}
\caption{Nonlinear least-squares fits (dashed lines) to the wind velocity profiles from simulations. The top panel shows the fits to the 1D simulations of \citet{Malfait2024} (solid lines) and the bottom panel shows the fit to the profile (diamonds) presented in \citet{Saladino19}. In both panels the gray shaded areas represent the radial range covered by the different orbits for different eccentricities.}
\label{fig:wind_malfait}
\end{center}
\end{figure}

To determine the $\beta$-law parameters that best describe the numerical results from simulations of eccentric orbits, we fit the wind velocity profiles for data presented by \citet{Malfait2024} and \citet{Saladino19}.

J. Malfait kindly provided us the 1D wind velocity profiles described in \citet{Malfait2024} corresponding to models with initial velocities $v_\mathrm{i}$ of 5, 10, and 20\,km s$^{-1}$. According to their Table 3, the corresponding terminal velocities for a molecular weight of 1.26 are 15.5, 17.7, and 24.8\,kms$^{-1}$. These profiles are plotted as solid lines in the top panel of Figure~\ref{fig:wind_malfait}. We performed a least-squares fit using the Marquardt-Levenberg algorithm in \texttt{gnuplot}\footnote{\url{http://www.gnuplot.info/}}. Equation~(\ref{eq:beta}) was used for the fits, with the $\beta$ exponent as the sole free parameter. The resulting best-fit curves, along with the derived $\beta$ values, are shown as dashed lines in the same panel. The panel also shows a gray area that highlights the radial range traveled by the accretor during its eccentric orbit with $e=0.5$.

We applied a similar approach to the wind profile presented in Figure 1 of \citet{Saladino19}, where a model with $v_\mathrm{i}=12$\,km s$^{-1}$ and $v_\infty=15.1$\,km s$^{-1}$ is analyzed. Our fit, shown in the bottom panel of Figure~\ref{fig:wind_malfait} with a dashed line, yielded $\beta=0.88$ but required a minor adjustment to the terminal velocity, $v_\infty=15.4$\,km s$^{-1}$. These parameters are employed to model the four eccentric cases discussed in \citet{SaladinoPols2019}, which correspond to simulations performed by the same group. The right panel also highlights the radial range covered during the different trajectories of the \citet{SaladinoPols2019}'s simulations.

\end{document}